\documentclass[11pt,a4paper,abstract=on]{scrartcl}
\usepackage[utf8]{inputenc}
\usepackage[english]{babel}
\usepackage[paper=a4paper,left=25mm,right=25mm,top=25mm,bottom=30mm]{geometry}
\usepackage{amsmath,amssymb,amsthm,mathrsfs,bbm,amsfonts,bm}
\usepackage{color,graphicx,cases,caption,enumerate}
\usepackage{float,multicol,authblk}
\usepackage[dvipsnames]{xcolor}
\usepackage{colortbl}
\usepackage{booktabs}% http://ctan.org/pkg/booktabs
\usepackage{multirow}% http://ctan.org/pkg/multirow
\usepackage{subcaption}
\usepackage{algorithmic}
\usepackage[linesnumbered,ruled]{algorithm2e}
\usepackage{setspace}
\usepackage{booktabs}
\usepackage{bbm}
\usepackage{capt-of}
\usepackage{array}
\usepackage[title]{appendix}
\newcolumntype{C}[1]{>{\centering\let\newline\\\arraybackslash\hspace{0pt}}m{#1}}
\usepackage{natbib}
\usepackage{placeins} % --> \FloatBarrier
\usepackage[breaklinks, colorlinks=true, citecolor=black, urlcolor=black, linkcolor=black]{hyperref}

\usepackage[stable]{footmisc}
\usepackage{comment}

\usepackage{tikz}
\usetikzlibrary{decorations.pathreplacing}
\usepackage{dsfont}

\DeclareMathOperator*{\argmax}{argmax}

\newif\ifclean
\cleantrue
\ifclean
    \long\def\gray#1{{}}
\else
    \long\def\gray#1{{\color{gray}{#1}\color{black}}}
\fi

\usepackage{bigstrut}

\title{Probabilistic intraday electricity price forecasting using generative machine learning}

\author[1]{Jieyu Chen}
\author[1,3,4]{Sebastian Lerch}
\author[1,3]{Melanie Schienle}
\author[2]{Tomasz Serafin}
\author[2]{Rafa{\l} Weron}
\affil[1]{Institute of Statistics, Karlsruhe Institute of Technology}
\affil[2]{Department of Operations Research and Business Intelligence, Wrocław University of Science and Technology}
\affil[3]{Heidelberg Institute for Theoretical Studies}
\affil[4]{Department of Mathematics and Computer Science, Marburg University}

\date{\today}

\begin{document}

\maketitle

\begin{abstract}
\noindent
% max 250 words
The growing importance of intraday electricity trading in Europe calls for improved price forecasting and tailored decision-support tools. 
In this paper, we propose a novel generative neural network model to generate probabilistic path forecasts for intraday electricity prices and use them to construct effective trading strategies for Germany's continuous-time intraday market.
Our method demonstrates competitive performance in terms of statistical evaluation metrics compared to two state-of-the-art statistical benchmark approaches. 
To further assess its economic value, we consider a realistic fixed-volume trading scenario and propose various strategies for placing market sell orders based on the path forecasts.
Among the different trading strategies, the price paths generated by our generative model lead to higher profit gains than the benchmark methods.
Our findings highlight the potential of generative machine learning tools in electricity price forecasting and underscore the importance of economic evaluation.
\end{abstract}

\section{Introduction}\label{sec:Introduction}

Since the introduction of competitive electricity markets in the 1990s, the day-ahead auction has played a central role in power trading \citep{may:tru:18, wer:14}.
However, the increasing use of renewable energy sources (RES) is gradually shifting market activity toward intraday (ID) trading.
Since 2015, trading volumes in the European ID markets operated by the European Power Exchange (EPEX) have increased by 300\%, while day-ahead volumes have risen by only 30\% \citep{epex:25}.

This trend is making its way into the electricity price forecasting (EPF) literature, albeit with some delay.
Of all Scopus-indexed publications from the years 2000-2009, only 5\% focused on predicting ID (or real-time) prices.\footnote{We used the Scopus query \texttt{TITLE((forecast* OR predict*) AND price*) AND TITLE-ABS-KEY("electric* market" OR "power market")} combined either with \texttt{AND TITLE-ABS-KEY("day-ahead" OR "spot" OR "next-day")} to identify DA-related or with \texttt{AND TITLE-ABS-KEY("intraday" OR "intra-day" OR "real-time")} to identify ID-related publications. Naturally, some of these papers concern both day-ahead and ID price forecasting.}
The share increased to 11\% in the next decade and then rapidly rose to 17\% in just the last five years.
One likely reason for the slower uptake is the diversity of market designs  \citep{gla:jos:pol:21}, making it difficult to compare the findings between studies.
For instance, North American real-time markets typically operate under a mandatory, security-constrained economic dispatch framework, whereas European markets often rely on voluntary ID auctions and/or continuous-time trading, which precedes the final settlement in the balancing market \citep{bac:kel:kra:23,cra:17,mac:uni:wer:23}.

The existing literature on price forecasting in ID electricity markets considers different perspectives.
Some studies aim to predict ID prices for the next day to take advantage of arbitrage opportunities \citep{mac:nit:wer:21}, to optimize the scheduling of a behind-the-meter storage system \citep{chi:z-d:zar:par:18}, or to manage the risk associated with trading \citep{kle:smi:not:23,jan:woj:22,bro:18}.
Others focus on very short-term forecasts, with lead times ranging from a few hours \citep{mon:r-r:f-j:con:16,uni:mar:wer:19,nar:zie:20JCM} to an hour or less before delivery \citep{bro:gil:22,bun:gia:kre:18}.
Many of these studies focus on probabilistic forecasts in the form of predictive distributions, which quantify predictive uncertainty and thus offer essential information for decision-making. 
In particular, multivariate probabilistic forecasts that capture temporal dependencies across different time stamps of ID price paths are of growing interest, as highlighted in the recent work by \citet{hirsch2025online}.

While traditional econometric models remain in use for EPF tasks \citep{jan:puc:23,mac:22:ORD,rus:kra:ber:kel:22}, they are increasingly being replaced by statistical learning methods \citep{nar:zie:20JCM,uni:mar:wer:19} and deep learning models \citep{oks:ugu:19,zha:wu:22,kle:smi:not:23,cra:etal:23}, which generally achieve superior predictive accuracy.
In recent years, deep neural networks have also gained traction in other high-volatility financial domains, such as stock markets \citep{chen2024deep, aleti2025intraday}.
However, to the best of our knowledge, with the exception of \citet{jan:ste:19} and \citet{hir:zie:24}, no neural network-based model has been proposed in the literature to predict marginal or joint multivariate distributions with temporal dependencies in a continuous-time ID electricity market.

Model inputs also vary, particularly in studies of European continuous-time ID electricity markets.
Most existing approaches rely on aggregate price indicators such as the ID3 index, which represents the volume-weighted average price of all transactions executed within the last three hours before delivery of a contract \citep{mac:22:ORD,uni:mar:wer:19,nar:zie:20JCM,rus:kra:ber:kel:22,cra:etal:23}.
While this aggregation offers a convenient summary of price evolution, it neglects the potential trading opportunities that arise from the RES generation updates \citep{kup:woz:23}.
Moreover, the timing of individual transactions plays a critical role in determining trading revenues \citep{ser:mar:wer:22,jan:ste:19}.
Therefore, the ability to simulate realistic ID price path trajectories is highly valuable for market participants.
Despite its practical importance, research on this topic remains scarce, with only a few notable contributions such as \citet{nar:zie:20}, \citet{ser:mar:wer:22}, and \citet{hir:zie:24} addressing this challenge.

To address these gaps, we propose a generative neural network model designed to predict multivariate distributions of the ID price path, capturing temporal dependencies to generate realistic price trajectories.
Our method is a data-driven, nonparametric approach, where the neural network directly outputs ID price path trajectories, incorporating information from historical price data and relevant exogenous input variables.
This approach builds on the conditional generative model (CGM) developed for multivariate probabilistic weather forecasting by \citet{che:etal:24}, which in turn extends earlier work of \citet{jan:ste:20} on multivariate prediction of day-ahead prices.
The CGM belongs to the class of scoring rule-based generative neural networks, where the model generates meaningful data from noise and is optimized using a loss function that measures the discrepancy between generated and real data.
Training the CGM involves optimizing a suitable multivariate proper scoring rule, e.g., the energy score, that quantifies the discrepancy between multivariate forecast samples (i.e., the price path trajectories) and a realization vector representing the temporal path of observed ID prices.
By conditioning on explanatory inputs, the model effectively captures nonlinear relationships for both marginal forecast distributions and temporal dependencies in the price paths, and integrates them into the output path trajectories.
Our CGM approach is in contrast to the commonly followed two-step framework for multivariate probabilistic forecasting, which proceeds by separately modeling the marginal distributions and the multivariate dependencies.
Such a two-step framework has been adopted in many disciplines, including EPF \citep{zie:wer:18} and weather forecasting based on ensemble post-processing \citep[e.g.,][]{schefzik_etal_2013_uncertainty,lerch_etal_2020_simulationbased,lakatos_etal_2023_comparison}. 

The specific application in electricity markets highlights the need for evaluating the performance of probabilistic forecasts from both statistical and economic perspectives.
Although many statistical measures have been proposed to assess the calibration and accuracy of univariate and multivariate probabilistic forecasts, these metrics typically do not directly correspond to the economic value obtained in real market scenarios.
The utilization of probabilistic multivariate forecasts for making optimal trading decisions and the economic evaluation of specific trading behaviors are thus of particular importance in this context.
Following previous research by \citet{ser:mar:wer:22}, we consider a simple trading scenario and propose several strategies to make optimal trading decisions based on multivariate probabilistic forecasts of electricity prices, and evaluate their performance based on an economic assessment of profit gains.

The remainder of this paper is structured as follows.
Section \ref{sec:data} provides a comprehensive description of the datasets used in this study.
In Section \ref{sec:methodology}, we present three approaches to probabilistic path forecasting of ID electricity prices, including the proposed CGM and two statistical benchmark methods.
In Section \ref{sec:Evaluation}, we describe the scoring rules utilized to evaluate the accuracy of path forecasts and introduce trading strategies applied for economic assessment in a case study.
Section \ref{sec:results} presents the results of both the statistical and economic evaluations, and discusses the practicality and effectiveness of these methodologies.
Finally, Section \ref{sec:conclusions} concludes the key findings of this research.
Python code for implementations of all forecasting methods is available online (\url{https://github.com/jieyu97/epf_cgm}).

\section{Data}\label{sec:data}

\begin{figure}
    \centering                   
    \includegraphics[width = 0.95\textwidth]{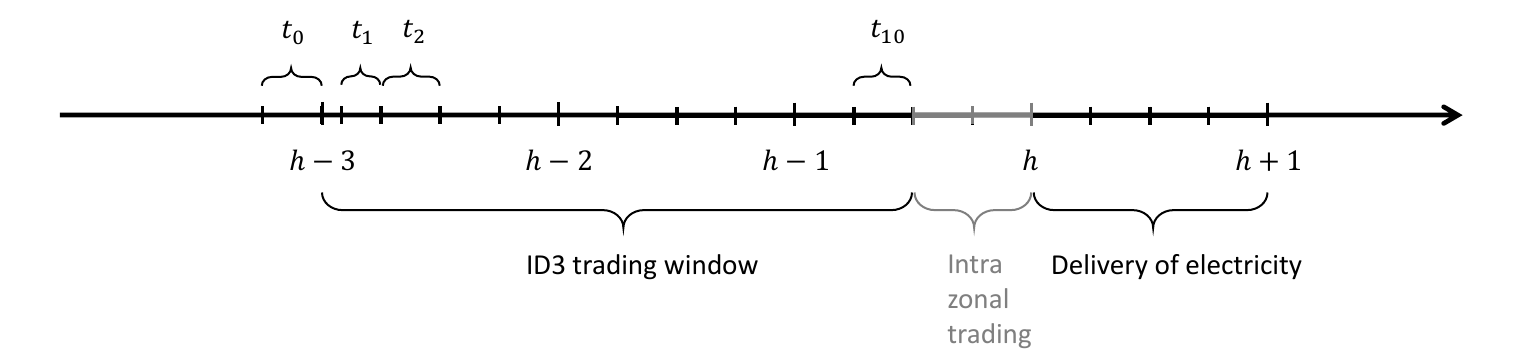}
    \caption{Timeline of the forecasting framework.
    Forecasts for ten 15-minute subperiods, denoted by $t_1, \ldots, t_{10}$, are generated three hours prior to delivery.
    The last 30 minutes before delivery, during which trading is restricted to within control zones, are excluded from the analysis.
    Note that the first subperiod, $t_1$, covers only 10 minutes, as the first five minutes are reserved for data collection and model execution.}
    \label{fig:timeline}
\end{figure}

The German ID electricity market offers both auction-based and continuous-time trading for hourly, half-hourly, and quarter-hourly products.
In this study, we focus exclusively on the continuous-time market for hourly delivery periods, which represents the most liquid segment \citep{epex:25, nar:zie:20JCM}.
Trading for these products begins at 16:00 on the day preceding delivery and ends 30 minutes prior to delivery, or five minutes prior within control zones.
Unlike auction-based mechanisms, prices in the continuous-time market evolve dynamically in real time as transactions occur between market participants, resembling the behavior of financial markets with a limit order book structure \citep{kup:woz:21}.

We consider ID price trajectories spanning the period from 15.06.2017 to 29.09.2019\footnote{The same dataset as in \citet{ser:mar:wer:22}.}, before the start of the crisis periods with COVID-19 and the Russian attack on Ukraine. 
Like \citet{ser:mar:wer:22}, we focus on the Volume Weighted Average Prices (VWAPs) of all transactions in the ten 15-minute subperiods, denoted as $t_1, t_2, ..., t_{10}$, ranging from three hours to 30 minutes before delivery, see Figure \ref{fig:timeline}.
The first subperiod, $t_1$, covers only 10 minutes, as the beginning five minutes are reserved for data collection and model execution. 
The last 30 minutes before delivery, $t_{11}$ and $t_{12}$, during which trading is restricted to within control zones, are excluded from the analysis.
The intraday VWAP path, denoted by $\bm{X}_{d,h} = (X_{d,h,t_j})_{j=1}^{10}$, at the ten subperiods $\{t_j\}_{j=1}^{10}$ for a specific hourly market at day $d$ and hour $h$, is the target to predict.
Our objective is to generate probabilistic multivariate forecasts in the form of path trajectories that capture the temporal dependencies across these subperiods.

In addition to the intraday VWAPs for 15-minute subperiods, six explanatory variables are available to be used as predictors for making path forecasts, including
\begin{itemize}
    \item the ID3 index $\texttt{ID3}_{d^*,h^*}$, which is defined as the VWAP of all transactions that took place in the last three hours before delivery of a given hourly product, and corresponds to the volume weighted average of the VWAPs over $\{t_j\}_{j=1}^{12}$;
    \item the day-ahead price $\texttt{DA}_{d^*,h^*}$, provided by the EPEX SPOT exchange\footnote{See \url{https://www.epexspot.com/en/indices}.};
    \item the real values of total load $L_{d^*,h^*}$ and its day-ahead forecasts $\hat{L}_{d^*,h^*}$, provided by the transmission system operator (TSO);
    \item the real values of wind generation $W_{d^*,h^*}$ and its day-ahead forecasts $\hat{W}_{d^*,h^*}$, provided by the TSO.
\end{itemize}
The indices $d^*$ and $h^*$ represent the day and the hour, respectively.
Multiple selected values $(d^*, h^*)$ are utilized to make path forecasts for the target hourly market at day $d$ and hour $h$.
All data series except the ID3 index are freely available from the ENTSO-E platform\footnote{See \url{https://transparency.entsoe.eu/}.}.
We assume that the actual values of the load and wind generation are available with a delay of less than 3 hours in real-time operation.

Like in \citet{ser:mar:wer:22}, the out-of-sample test period comprises the last 200 days (from 13.03.2019 to 29.09.2019).
The preceding data is used for model training and generating path trajectories using different approaches.
All predictors and target variables are normalized to ensure more stable and efficient training, where different standardization schemes are applied and will be introduced separately for each approach in the following.

\section{Methods}\label{sec:methodology}

\begin{figure}
\centering
    \includegraphics[width = 0.95\textwidth]{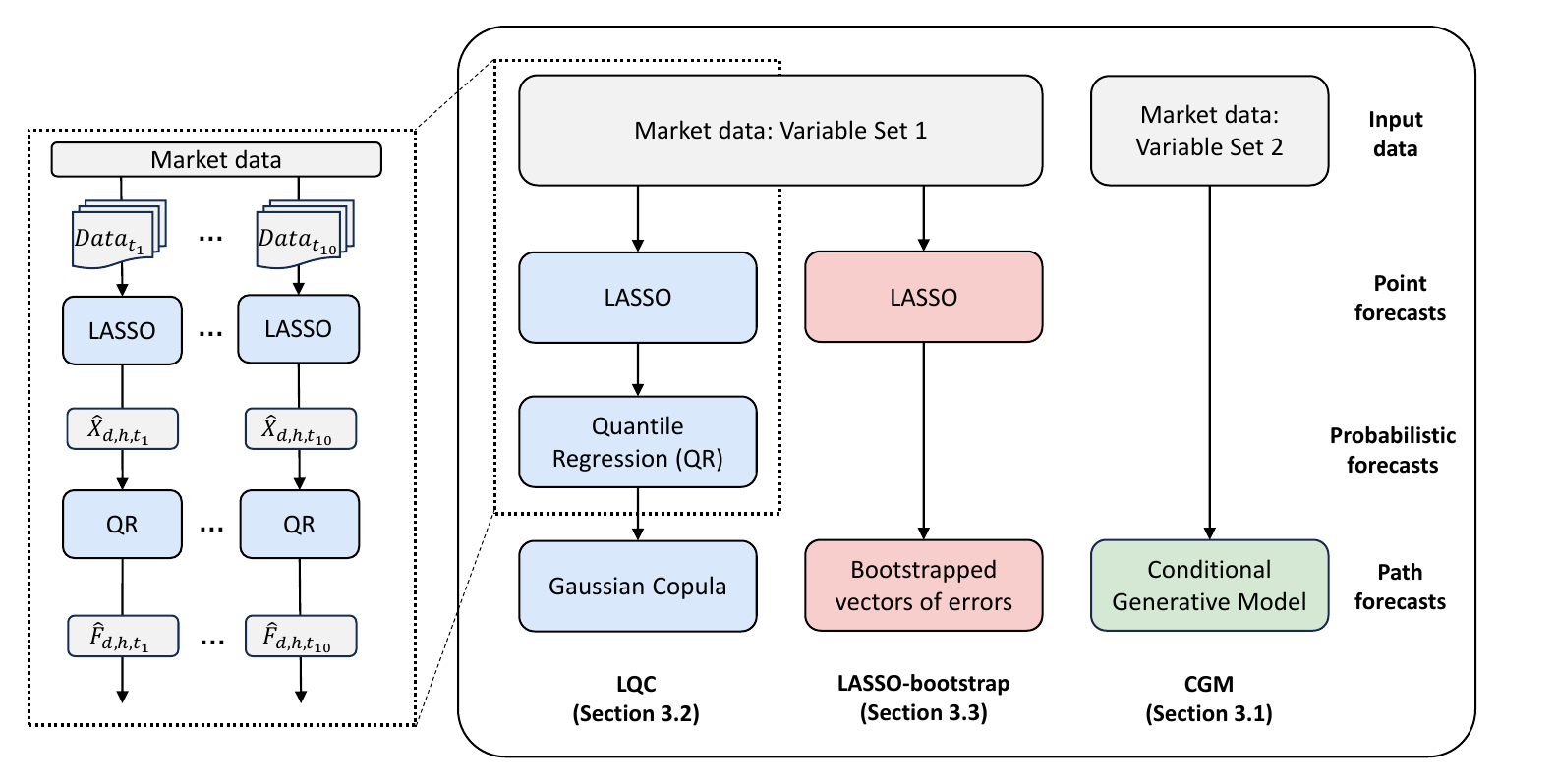}
    \caption{Schematic overview of the three approaches to multivariate probabilistic time-series forecasting.}
    \label{fig:epf_3methods}
\end{figure}

This section introduces three approaches to multivariate probabilistic time-series forecasting for generating ID electricity price trajectories across multiple subperiods.
These include the proposed generative machine learning method based on a conditional generative model (CGM), as well as two state-of-the-art statistical benchmark methods originally introduced by \citet{ser:mar:wer:22}.
A schematic overview of the three approaches is presented in Figure \ref{fig:epf_3methods}.
Many multivariate time-series forecasting methods in the EPF literature adopt a two-step framework, wherein marginal predictive distributions are estimated first, followed by a separate modeling of temporal dependencies \citep{zie:wer:18}.
This structure is also employed by the two benchmark methods considered in this study.
In contrast, the proposed CGM approach integrates both steps into a unified framework, directly generating multivariate forecast trajectories that inherently capture temporal dependencies.

\subsection{The proposed conditional generative model}\label{ssec:CGM}

We propose a novel approach to directly produce multivariate time-series forecasts in the form of path trajectories using generative machine learning.
This approach builds on the framework developed by \citet{che:etal:24} in the context of multivariate post-processing of ensemble weather forecasts.
Our conditional generative model (CGM) is a nonparametric approach which does not require parametric assumptions on the marginal distribution or the multivariate dependence structure.
This is achieved by utilizing an implicit generative neural network that parametrizes the stochastic process of generating meaningful data from noise, and directly yields simulated ID price path trajectories as output.
Incorporating information from the available exogenous predictors as inputs enables the CGM to learn complex and nonlinear relationships within the data.
The CGM is trained by minimizing the energy score, which will be introduced in Section \ref{sssec:ES}, as a loss function that measures the discrepancy between the generated path trajectories and the observed multivariate ID price path.
For a more detailed description of the mathematical background of generative models and the CGM, we refer to \citet{che:etal:24}.

From a conceptual standpoint, the CGM offers a key advantage over traditional two-step frameworks in multivariate probabilistic forecasting by streamlining the training process.
Unlike traditional approaches that separately estimate marginal forecasts and model multivariate dependencies via copulas in a post hoc manner, the CGM directly generates multivariate probabilistic forecasts, thereby simplifying model training and reducing potential sources of error propagation.
Moreover, the flexibility of incorporating exogenous predictors in the CGM allows additional information to be used in modeling multivariate dependencies, whereas copula-based methods typically rely solely on historical target variables.
Another strength of the CGM lies in its flexibility with respect to the training loss function, where the energy score is not the only viable choice.
While \citet{pacchiardi_etal_2024_probabilistic} discussed the use of other multivariate proper scoring rules in a similar setting, we explore the use of a custom loss function tailored to the needs of the economic evaluation, see Section \ref{ssec:cgm:newloss}.

\subsubsection{Model architecture}

\begin{figure}
\centering
    \includegraphics[width = \textwidth]{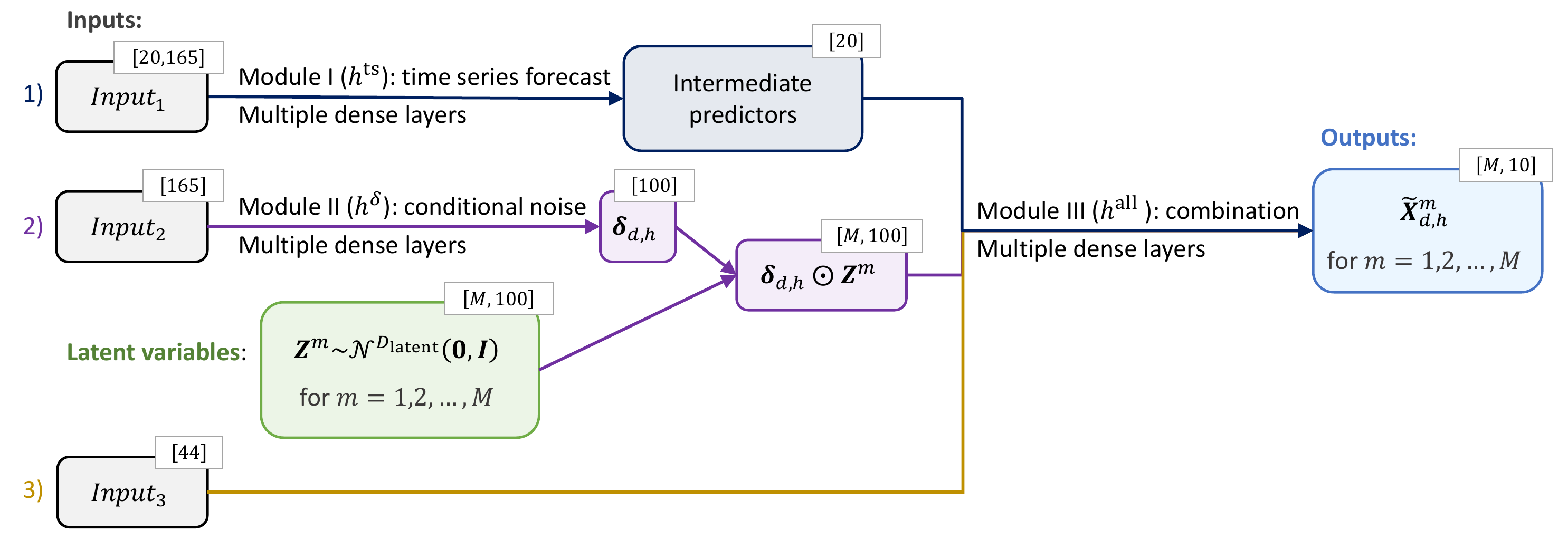}
    \caption{Schematic illustration of the conditional generative model (CGM) used to generate $M$ path trajectories of the multivariate ID price forecast for a given hourly market at day $d$ and hour $h$.
    The dimensions of the tensors at each module are indicated in the small boxes, with the batch size omitted.}
    \label{fig:cgm}
\end{figure}

Figure \ref{fig:cgm} provides a schematic illustration of our CGM.
The output of the CGM is a set of 10-dimensional vectors,
$$\tilde{\bm{X}}^m_{d, h} = (\tilde{X}^m_{d, h, t_1}, \ldots, \tilde{X}^m_{d, h, t_{10}}),$$
representing sample paths of ID prices over the 10 subperiods from the underlying multivariate forecast distribution for a target hourly market at day $d$ and hour $h$.
The model comprises three components to efficiently incorporate relevant exogenous predictors in different segments and to propagate relevant uncertainty information to the generated ID paths by transforming the input noise of the generative model.
This design results in three separate input modules, with the corresponding parts represented in different colors in the schematic illustration.

The first module of the model, denoted by $h^{\text{ts}}$, aims at generating intermediate predictions as latent information for the subsequent parts.
It is designed to mimic deterministic time-series forecasting and utilizes a fully connected feed-forward neural network.
The input for this module, denoted by $\textit{Input}_1$, consists of 20 predictor variables, including the six exogenous variables introduced in Section \ref{sec:data}, the VWAP at 12 subperiods before delivery (from $t_1$ to $t_{12}$) along with their standard deviations, and the VWAP of the last subperiod preceding $t_1$.
This VWAP, which is denoted by $X_{d^*,h^*,t_0}$, corresponds to the period from three hours 15 minutes to three hours before the delivery of a target hourly market at day $d^*$ and hour $h^*$.
For all 20 input variables, we use a window of historical data ranging from one week to four hours before the delivery time.
For the ID price-related predictors, the data corresponds to historical hourly markets.
The full list of inputs for this first module thus is
% \begin{align*}
%     \textit{Input}_1 &= \Big\{ \texttt{ID3}_{d,h-i}, \texttt{DA}_{d,h-i}, L_{d,h-i}, \hat{L}_{d,h-i}, W_{d,h-i}, \\
%     & \hat{W}_{d,h-i}, \{X_{d,h-i,t_j}\}_{j=0}^{12}, \sigma \big( \{X_{d,h-i,t_j}\}_{j=1}^{12} \big) \Big\}_{i=4}^{168}.
% \end{align*}
\begin{equation*}
    \textit{Input}_1 = \Big\{ \texttt{ID3}_{d,h-i}, \texttt{DA}_{d,h-i}, L_{d,h-i}, \hat{L}_{d,h-i}, W_{d,h-i}, \hat{W}_{d,h-i}, \{X_{d,h-i,t_j}\}_{j=0}^{12}, \sigma \big( \{X_{d,h-i,t_j}\}_{j=1}^{12} \big) \Big\}_{i=4}^{168}.
\end{equation*}

The second module, denoted by $h^\delta$, is the core of the generative model based on which it learns to produce meaningful noise estimates conditional on the available input data.
We generate latent noise variables by sampling from a standard multivariate Gaussian distribution, which is a common choice in generative models.
The dimensionality of the latent variables $D_{\text{latent}}$ is a hyperparameter of the model that controls the complexity of randomness for each sample and needs to be determined through hyperparameter tuning. 
We use $\bm{Z}^m$ to denote a single sample of the noise vector from which we eventually obtain the corresponding output sample $\tilde{\bm{X}}^m_{d,h}$ as the final output of our generative model.
By repeatedly generating samples from the noise distribution and propagating them through the generative model, we obtain a multivariate probabilistic forecast in the form of path samples as output.
The number of noise samples we draw during training (and inference) determines the number of output path trajectories, and thus enables the generation of arbitrarily many sample trajectories.

The scale of the generated latent noise $\bm{Z}^m$ is adjusted by incorporating uncertainty information from the second part of the available inputs, denoted by $\textit{Input}_2$, which utilizes the standard deviation predictor in $\textit{Input}_1$.
We refer to the output of this scale adjustment as conditional noise.
A fully connected feed-forward neural network is employed to learn the adjusted scales $\bm{\delta}_{d, h}$ for all latent variables, and the conditional noise is obtained via
\begin{equation*}
    h^\delta\big( \textit{Input}_2 \big) \odot \bm{Z}_m, \; \bm{Z}_m \sim \mathcal{N}^{D_{\text{latent}}}(\bm{0}, \bm{I}); \quad \text{with } \, \textit{Input}_2 = \Big\{ \sigma \big( \{X_{d,h-i,t_j}\}_{j=1}^{12} \big) \Big\}_{i=4}^{168}.
\end{equation*}

The third and final module, denoted by $h^\text{all}$, further incorporates more recent historical information available within the four hours before delivery, and integrates the intermediate predictions and conditional noise from the previous two modules to generate sample trajectories. 
The inputs for this module, denoted by $\textit{Input}_3$, contain the 20 variables from $\textit{Input}_1$, but only for the specific values at $(d^*, h^*)$ corresponding to four hours before delivery of the target hourly market. 
Additionally, it incorporates four variables, including the day-ahead price, the last VWAP, and the day-ahead forecasts of wind generation and load, available from three hours before up to the delivery time.
While the observed ID price path of previous hourly markets within three hours before the target delivery is not fully available at the time of forecasting, partial paths are accessible and can provide valuable insights into the latest real-time ID prices.
Therefore, we incorporate these available ID prices, specifically $\{X_{d,h-2,t_j}\}_{j=9}^{12}$ and $\{X_{d,h-3,t_j}\}_{j=5}^{12}$.
We also incorporate time dummy variables to convey the time information of the target hourly market, including both sine and cosine transforms of "the day of a year" $d$, and "the hour of a day" $h$.
The weekday information, ranging from one to seven, is treated as a separate input component.
This information is integrated with the other inputs after being processed through an embedding layer that converts categorical integer values into two-dimensional vectors, following related work in probabilistic weather forecasting \citet{ras:ler:18}, which in turn is based on widely used embedding techniques in natural language processing.
The complete list of $\textit{Input}_3$ is
\begin{align*}
    \textit{Input}_3 =& \, \Big\{ \textit{Input}_1(i=4), \big\{ \texttt{DA}_{d,h-i}, \hat{L}_{d,h-i}, \hat{W}_{d,h-i}, X_{d,h-i,t_0} \big\}_{i=0}^{3}, \\
    & \quad \{X_{d,h-2,t_j}\}_{j=9}^{12}, \{X_{d,h-3,t_j}\}_{j=5}^{12}, \{\text{time indicators}\} \Big\}.
\end{align*}
As final output of the CGM, we thus obtain samples of path trajectories via
\begin{equation*}
    \tilde{\bm{X}}^m_{d, h} = h^\text{all} \big( h^{\text{ts}}(\textit{Input}_1), h^\delta\big( \textit{Input}_2 \big) \odot \bm{z}_m, \textit{Input}_3 \big)
\end{equation*}
with $m = 1,2,\ldots$, by repeatedly generating samples from the latent noise distribution.

\subsubsection{Implementation details}

The CGM is trained by minimizing the empirical energy score, see Section \ref{sssec:ES}.
To reduce the randomness inherent in neural network training, we generate an ensemble of 10 CGMs by training separate models with identical hyperparameters on the same data, but with different random seeds.
This strategy has proven effective in improving robustness and overall forecast quality, and is competitive with other ensemble generation mechanisms for neural network-based forecasting models \citep{schulz_etal_2024_aggregating}.
Each ensemble run generates 1\,000 output samples, and the combined output of all ensemble runs yields a total of 10\,000 forecast path trajectories as the final outcome of the model.
For a detailed investigation of strategies to generate ensembles of CGMs, see \citet{che:etal:24}.

The hyperparameters determining the structure of the CGM, including the number of layers, nodes, activation functions in each layer, and the number of latent variables, need to be determined through hyperparameter tuning.
The hyperparameters were determined based on a combination of exploratory experiments and an additional grid search.
The three model components have different hyperparameter configurations, and the overall framework consists of 100 latent variables for the noise component, 10 dense layers, with the ELU activation function \citep{elu2015} used for most layers.
For a complete list of hyperparameter choices used by the CGM, we refer to the Python code accompanying this work.
The model is trained using stochastic gradient descent optimization with the Adam optimizer \citep{kingma2014adam} with a learning rate of $1 \times 10^{-4}$, a batch size of 1\,024, and an early stopping criterion with a patience of 10 epochs to avoid overfitting.

The CGM is trained over a fixed period of 630 days (22.06.2017--13.03.2019), with 20 percent of the data randomly selected as the validation set.
The training period begins one week after the first date in the original dataset to ensure historical input variables are available.
All input variables are normalized by subtracting the mean and dividing by the standard deviation of the data over the training period.
Preliminary experiments indicated no improvements, and sometimes worse performance, when training the CGM with a sliding window, likely due to higher variability in the training data.
Therefore, in contrast to the statistical benchmark methods, we do not employ sliding window training for the CGM, even though the comparatively low computational cost would have made this technically possible, as the training process only takes a few minutes on multiple CPUs.
That said, rolling window training may still offer advantages for different datasets and contexts, and thus may be worth considering to enable the model to better adapt to structural changes in the data over time.

\subsection{LQC benchmark}\label{ssec:LQC}

The LQC approach \citep{ser:mar:wer:22} comprises three components: a deterministic point prediction model, a transformation of those point predictions to probabilistic forecasts, and a restoration of temporal dependencies. 
The specific methods applied in those components lend the LQC approach its name: The point predictions are obtained via (L)ASSO-estimated (auto)regression \cite[also known as the LEAR model;][]{lag:mar:des:wer:21}, and are converted to probabilistic predictions via (Q)uantile regression \cite[as in the quantile regression averaging approach proposed by][]{now:wer:15}. Finally, a Gaussian (C)opula is employed for modeling temporal dependencies \cite[as suggested in][]{pin:etal:09}.

In the first step, point predictions are made using the LEAR model, utilizing 102 inputs (or regressors) derived from the six explanatory variables introduced in Section \ref{sec:data}:
\begin{itemize}
\item $\{\texttt{ID3}_{d, h-i}\}_{i=4}^{24}$, i.e., the most recent 21 historical ID3 index values available at the time of prediction;
\item $\{\texttt{DA}_{d, h-i}\}_{i=0}^{24}$, i.e., 25 day-ahead prices available within one day before delivery;
\item $\{\hat{W}_{d, h-i}, \hat{L}_{d, h-i}\}_{i=0}^{24}$, i.e., 25 hourly values of day-ahead wind generation and load forecasts available within one day before delivery;
\item $\{W_{d, h-4}, W_{d, h-24}, L_{d, h-4}, L_{d, h-24}\}$, i.e., the actual wind power production and observed load for the last observed hour (4 hours before delivery) and 24 hours ago;
\item $\{X_{d,h,t_0}\}$, i.e., the last VWAP spanning the transaction period from 3 hours 15 minutes to 3 hours before delivery;
\end{itemize}
where day $d$, hour $h$ represent the delivery time of the target hourly market.
Separate models are constructed for each of the 10 subperiods $j=1,2,\ldots,10$, and the least absolute shrinkage and selection operator \citep[LASSO;][]{tib:96} is used to remove redundant features.

Following \citet{tsc:etal:22} and \citet{zie:wer:18}, we transform the inputs by applying the \emph{area hyperbolic sine}\footnote{The area (inverse) hyperbolic sine can be computed by $\mathrm{arsinh}(x) = \ln\left(x + \sqrt{x^2 + 1}\right)$.}.
As suggested by \citet{uni:wer:zie:18}, each input series is first independently normalized by subtracting the in-sample median and dividing by the in-sample median absolute deviation, adjusted by the 75th percentile of the standard normal distribution.
Once the point prediction $\hat{X}_{d,h,t_j}$ for each subperiod $j$ is generated, the transformation and normalization are inverted.

Based on the point forecast processed separately for different subperiods, we use quantile regression \cite[QR;][]{koe:05} in the next step to compute empirical forecasts in the form of 99 percentiles of the predictive distribution $\hat{F}_{d,h,t_j}$ at each margin, i.e., for each subperiod $t_j$ before delivery at day $d$ and hour $h$.
The LASSO and QR steps result in probabilistic forecasts of the marginal distribution and thus constitute the first part of the two-step framework for multivariate forecasting.
The 99 percentiles are linearly interpolated, with linear extrapolation applied to the minimum and maximum prices for the extreme values, to allow for drawing arbitrarily many quantiles in the subsequent step.

In the final step, multivariate path trajectories of ID prices across multiple subperiods are generated based on the predicted quantiles, with temporal dependencies between subperiods modeled using a Gaussian copula, as presented in \citet{ser:mar:wer:22}.
The probabilistic path forecasts consisting of $M$ trajectories 
$$
\Big\{ \tilde{\bm{X}}^m_{d, h} = (\tilde{X}^m_{d, h, t_1}, \ldots, \tilde{X}^m_{d, h, t_{10}}) \Big\}_{m=1}^M
$$
of the target ID prices 
$$
\bm{X}_{d, h} = (X_{d, h, t_1}, \ldots, X_{d, h, t_{10}})
$$
are derived from $M$ random samples $\big\{\bm{Z}^m_{d, h} = (Z^m_{d, h, t_1}, \ldots, Z^m_{d, h, t_{10}})\big\}_{m=1}^M \sim \mathcal{N}(\mathbf{0}, \mathbf{\Sigma}_{d, h})$ from a multivariate Gaussian distribution, i.e.,
\begin{equation*}
    \tilde{X}^m_{d, h, t_j} = \hat{F}^{-1}_{d, h, t_j} \big( \Phi \Big( Z^m_{d, h, t_j} \big) \Big), \quad \text{for } j = 1, \ldots, 10,
\end{equation*}
where  $\hat{F}^{-1}_{d, h, t_j}$ denotes the inverse transformation of the marginal forecast CDF $\hat{F}_{d, h, t_j}$, and $\Phi$ represents the standard Gaussian CDF.
The covariance matrix $\mathbf{\Sigma}_{d, h}$ is estimated based on the transformed historical ID prices from a preceding calibration window $\mathcal{C} =  [d-120, d)$,
\begin{equation*}
    \mathbf{\Sigma}_{d, h} = \mathrm{cov}\left( \big\{ \hat{\bm{Z}}_{d^*, h} = (\hat{Z}_{d^*, h, t_1}, \ldots, \hat{Z}_{d^*, h, t_{10}}) \big\}_{d^* \in \mathcal{C}} \right), \quad \text{with }\, \hat{Z}_{d^*, h, t_j} = \Phi^{-1} \big( \hat{F}_{d^*, h, t_j} (X_{d^*, h, t_j}) \big).
\end{equation*}

For all three steps of the LQC approach, a rolling window scheme is employed.
Each day, the calibration windows are moved forward by one day to produce the next day's forecasts, with different window sizes used for each step.
We first use LASSO-estimated regression fitted to data from a 396-day calibration window (sliding window initially starting from 16.06.2017) to compute point predictions, then apply QR with parameters estimated using a 120-day calibration window (sliding window initially starting from 16.07.2018).
Once computed, the predictive distributions are converted into path forecasts using a Gaussian copula fitted over a 120-day calibration window (sliding window initially starting from 13.11.2018).

\subsection{LASSO bootstrap benchmark}\label{ssec:LASSO bootstrap}

The LASSO bootstrap approach uses the same point predictions from the LEAR model as the LQC approach.
These point predictions serve as the basis for obtaining probabilistic price path forecasts without the need to compute predictive distributions, utilizing a bootstrapping method.
Thereby, vectors of historical point forecast errors are sampled to incorporate temporal dependencies based on past obseervations.

To obtain a multivariate path trajectory $\tilde{\bm{X}}^m_{d,h}$ of ID price forecast for the delivery at day $d$ and hour $h$, we first compute vectors of past point forecast errors from a preceding calibration window, i.e.,
\[
\bm{\varepsilon}_{d^*,h} = \hat{\bm{X}}_{d^*,h} - \bm{X}_{d^*,h}, \quad \text{with } d^* \in [d - 240, d),
\]
and proceed by adding bootstrapped error vectors to the point predictions for the target path,
$$ 
\tilde{\bm{X}}^m_{d,h} = \hat{\bm{X}}_{d,h} + \bm{\varepsilon}^m_{d^*,h}, \quad \text{with } \bm{\varepsilon}^m_{d^*,h} \in \{\bm{\varepsilon}_{d^*,h}\}_{d^* \in [d - 240, d)}, \quad \text{for } m = 1, \ldots, M,
$$ 
where $\hat{\bm{X}}_{d,h} = (\hat{X}_{d,h,t_1}, \ldots, \hat{X}_{d,h,t_{10}})$ are the point predictions for all subperiods.

The LASSO bootstrap approach also employs a rolling window scheme.
The first step is the same as the LASSO step in the LQC approach, where we fit a LASSO-estimated regression model using data from a 396-day calibration window. 
In the next step, randomly sampled historical error vectors from a 240-day calibration window (sliding window initially starting from 16.07.2018) are added to the point predictions to obtain multivariate probabilistic forecasts of ID prices.

\section{Statistical and economic evaluation methods}\label{sec:Evaluation}

We here introduce various evaluation metrics that will be used in Section \ref{sec:results} to compare the CGM against the two statistical benchmarks.
We present widely used statistical metrics for probabilistic forecasts that account for prediction uncertainty, and propose economic evaluation methods based on trading strategies on top.
These are motivated from a practical perspective where a manager has to make a decision, and different evaluation metrics may point to different suggested actions \citep{kol:20}. 
At the same time, the optimal choice will be affected by the decision maker's preferences, e.g., regarding profit maximization or risk reduction. 
Statistical evaluation alone thus does not provide the necessary information, as there is no clear and obvious relationship between scoring metrics and the expected outcome of economic decisions.
This makes it unclear whether higher accuracy in terms of statistical evaluation metrics translates into better economic results in practice \citep{mac:uni:wer:23,yar:pet:21}. 
To address this, we consider a range of trading strategies based on the generated ID price path forecasts, which will be introduced in Section \ref{ssec:Economic:evaluation}, and evaluate different methods in a case study involving a fixed-volume scenario.

\subsection{Statistical evaluation}\label{ssec:Statistical:evaluation}

Since probabilistic forecasts capture prediction uncertainty, respective statistical evaluation metrics should also take uncertainty information into account.
The widely accepted standard tools for probabilistic forecast evaluation are proper scoring rules \citep{gne:raf:07}, which simultaneously assess calibration and sharpness of predictive distributions. 
In a nutshell, a scoring rule $S(F,x)$ assigns a numerical score to a pair of a forecast distribution $F$ and a realizing observation $x$.
It is called proper, if the true distribution of the observation achieves the best (i.e., lowest) possible score in expectation, i.e., $\mathbb{E}_{X\sim G} S(G,X) \leq \mathbb{E}_{X\sim G} S(F,X)$ for all pairs of forecast distributions $F, G$ from a suitably chosen class of probability distributions.
For details, we refer to \citet{gne:raf:07}, available software implementations \citep[e.g.,][]{jordan_etal_2019_evaluating}, and the wide variety of research in statistics and application disciplines, including, e.g., \citet{lauret_etal_2019_verification} with a focus on energy forecasting.

The continuous ranked probability score (CRPS), proposed by \citet{mat:win:76}, is a proper scoring rule widely used for evaluating univariate probabilistic forecasts.
Given marginal forecast CDF $\hat{F}_{d, h, t_j}$ and the real price $X_{d, h, t_j}$ at subperiod $t_j$ for hourly market day $d$ and hour $h$, the CRPS is defined as
\begin{equation*}
    \text{CRPS}_{d, h, t_j}(\hat{F}_{d, h, t_j}, X_{d, h, t_j}) = \int_{-\infty}^{\infty} \big( \hat{F}_{d, h, t_j}(z) - \mathbb{I} \{z \geq X_{d, h, t_j}\} \big)^2 dz,
\end{equation*}
where $\mathbb{I}$ denotes the indicator function.
Based on empirical samples $\{\tilde{X}_{d, h, t_j}^m\}_{m=1}^M$ from the predictive distribution, it can be formulated as 
\begin{equation*}
    \text{CRPS}_{d, h, t_j} = \frac{1}{M} \sum_{m=1}^M \big| \tilde{X}^m_{d, h, t_j} - X_{d, h, t_j} \big| - \frac{1}{2M^2} \sum_{m=1}^M \sum_{n=1}^M \big| \tilde{X}^m_{d, h, t_j} - \tilde{X}^n_{d, h, t_j} \big|.
\end{equation*}
The CRPS is negatively oriented and equals zero for a forecast that perfectly matches the observed distribution.
In the special case where the forecast is a deterministic point prediction, the CRPS reduces to the mean absolute error.

The direct generalization of CRPS to multivariate forecasts is the energy score, which will be introduced below.
In addition to the energy score, several proper scoring rules have been proposed for evaluating multivariate probabilistic forecasts.
However, all of them come with certain shortcomings in terms of sensitivity to certain types of misspecifications of the multivariate forecast distribution \citep{sch:ham:15,ale:cou:han:men:22}.
A comprehensive understanding of contributions to various types of misspecifications to the behavior of multivariate proper scoring rules remains an open question and subject of current research, see the discussion in \citet{che:etal:24} and references therein.
We here use three popular multivariate proper scoring rules: the energy score (ES), the Dawid-Sebastiani score (DSS) and the variogram score (VS).

\subsubsection{Energy score}\label{sssec:ES}

The energy score \citep[ES;][]{gne:raf:07} is given by
\begin{equation*}
    \text{ES}_{d,h} = \frac{1}{M} \sum_{m=1}^{M} \left\| \tilde{\bm{X}}^{m}_{d,h} - \bm{X}_{d,h} \right\|_{2} -\frac{1}{M(M-1)}\sum_{m=1}^{M-1} \sum_{n=m+1}^{M} \left\| \tilde{\bm{X}}^{m}_{d,h} - \tilde{\bm{X}}^{n}_{d,h} \right\|_{2},
\end{equation*}
where $\tilde{\bm{X}}^{m}_{d,h} = \big(\tilde{X}^{m}_{d,h,t_1},\ldots,\tilde{X}^{m}_{d,h,t_{10}} \big) $ is the $m$-th multivariate realization of ID price path forecast for day $d$ and hour $h$, $\bm{X}_{d,h}$ is the corresponding observed ID price path, and $M$ is the number of generated path trajectories. 
A number of studies have noted that the ES lacks sensitivity to misspecifications of the dependence structure \citep[e.g.,][]{pin:gir:12,ale:cou:han:men:22}.

\subsubsection{Dawid-Sebastiani score}\label{sssec:DSS}

The Dawid-Sebastiani score \citep[DSS;][]{daw:seb:99} is estimated based on the mean vector and covariance matrix of the predictive distribution
\begin{equation*}
    \text{DSS}_{d,h} = \log\left(\mathrm{det}\left(\mathbf{S}_{d,h}\right)\right) + \mathbf{K}^{T}\mathbf{S}^{-1}_{d,h}\mathbf{K},
\end{equation*}
where in our case $\mathbf{K}_{d,h} = \left(K_{d,h,t_{1}}, \ldots,K_{d,h,t_{10}} \right)$ is a vector of 10 differences,
\begin{equation*}
   K_{d,h,t_{j}} =  X_{d,h,t_{j}} - \frac{1}{M}\sum_{m=1}^M\tilde{X}_{d,h,t_{j}}^m
\end{equation*}
and $\mathbf{S}_{d,h}$ is the covariance matrix estimated from the simulated scenarios.
The DSS corresponds to the logarithmic score for multivariate Gaussian predictive distributions and is a proper scoring rule for a broad class of probability distributions. 

In addition to shortcomings that have been noted in cases where forecast accuracy is moderate \citep{wil:20}, a major limitation of this score is the potential numerical issue when inverting the covariance matrix if the sample size is small relative to the number of ensemble members  \citep{sch:ham:15}.

\subsubsection{Variogram score}\label{sssec:VS}

The variogram score \citep[VS;][]{sch:ham:15} is given by
\begin{equation*}
    \text{VS}_{d,h} = \sum_{i,j=1}^{10}
    w_{i,j} \left( \left|X_{d,h,t_i} - X_{d,h,t_j} \right|^{p} - \frac{1}{M} \sum_{m=1}^{M} \left|\tilde{X}^{m}_{d,h,t_i} - \tilde{X}^{m}_{d,h,t_j}\right|^{p} \right)^2,
\end{equation*}
where $p$ is the order of the VS, and $w_{i,j}$ is an optional weight parameter. We here consider the unweighted version with $w_{i,j} = \frac{1}{100}$.
It has been argued that the VS tends to be more discriminative than the ES and DSS when the correlation structure of ensemble forecasts is misspecified \citep{sch:ham:15}.
The order $p$ needs to be chosen by the user, with \citet{ale:cou:han:men:22} noting that the VS with $p=0.5$ has a superior discriminative ability when dealing with relatively accurate forecasts, whereas $p=1$ should be used in cases with moderate prediction accuracy.

\subsection{Economic evaluation}\label{ssec:Economic:evaluation}

To evaluate the generated path forecasts from an economic perspective, we consider a range of trading strategies for the fixed-volume scenario introduced by \citet{ser:mar:wer:22} in the continuous-time ID market.
The fixed-volume scenario assumes that an energy producer owning intermittent RES sells the surplus of 1~MWh of electricity in each hour of the day.
A similar setup has been considered by \citet{kat:zie:18} and \citet{jan:puc:23}, among others.
We make the standard assumption that the impact of our trades on the ID prices is negligible and ignore the transaction costs.
The decision problem can then be treated as finding the optimal time to enter the market for selling the fixed amount of electricity for each individual hourly delivery period.

In the following, we present two classes of strategies that rely on multivariate path forecasts for the fixed-volume scenario, where one is based directly on the multivariate trajectories and the other utilizes prediction bands derived from the path forecasts.
In addition, we describe the naive benchmark strategies and introduce a crystal ball (or orcale) benchmark to evaluate the realized trading potential when using multivariate price forecasts of the benchmark models and the proposed CGM.

\subsubsection{Trading strategies based on probabilistic forecasts} \label{ssec:trading:band}\label{ssec:trading:path}

\subsubsection*{Majority vote strategy}

Given a single path forecast in the form of a trajectory of ID price across multiple subperiods, the most intuitive and simple approach to determining the optimal time for selling the fixed amount of electricity is to simply use the subperiod when the predicted path trajectory reaches its maximum price.
Based on the collection of $M$ generated path trajectories which are obtained as outputs of the different forecasting methods, we use a majority-vote strategy to identify the most frequent subperiod with the maximum price.
The optimal time for entering the market using the majority-vote strategy for $M$ path trajectories $\big\{ \tilde{\bm{X}}^m_{d,h} = (\tilde{X}^m_{d,h,t_1}, \ldots, \tilde{X}^m_{d,h,t_{10}}) \big\}_{m=1}^M$ is then given by
\begin{equation}\label{eq:majority-vote}
    J_{d,h} = \operatorname{mode} \left( \left\{ \argmax_{j \in \{1,\ldots,10\}} \tilde{X}^m_{d,h,t_j} \right\}_{m=1}^M \right),
\end{equation}
where $J_{d,h}$ is the index of the optimal subperiod for selling the fixed amount of electricity.

\subsubsection*{Prediction band-based strategy}

In addition to selecting the optimal time for entering the market directly from the simulated trajectories of future price paths, we further explore strategies based on prediction bands derived from the collection of path forecasts, which were first proposed in \citet{ser:mar:wer:22}.

Prediction bands, unlike a set of prediction intervals, account for the temporal dependence in the evolution of predicted prices over time.
Each prediction band (upper or lower) is defined by the simultaneous coverage probability (SCP), which represents the probability that the entire price trajectory lies below ($\rightarrow$ upper) or above ($\rightarrow$ lower) the band.
More formally, the SCP for the upper prediction band $\bm{B}^U_{d,h,t_j}\in\mathbb{R}^{10}$ is given by
\begin{equation}
	\mathbb{P}\left( X_{d,h,t_j} \leq \bm{B}^U_{d,h,t_j}, \forall_{j} \right) = \text{SCP}, \nonumber
\end{equation}
and for the lower $\bm{B}^L_{d,h,t_j}$ by
\begin{equation}
	\mathbb{P}\left( \bm{B}^L_{d,h,t_j} \leq X_{d,h,t_j} , \forall_{j} \right) = \text{SCP}. \nonumber
\end{equation}

The algorithm used to construct prediction bands is based on \citet{sta:07}.
To satisfy the simultaneous coverage property, which requires that predicted price paths remain within the prediction band at all time points, the procedure involves filtering out simulated trajectories that contain extreme values.
Specifically, trajectories with maximum values (for the upper band) or minimum values (for the lower band) at any time subperiod are iteratively removed until only a fraction corresponding to the desired SCP level remains.
The prediction band is then constructed by taking the pointwise maximum (for the upper band) and minimum (for the lower band) at each subperiod across the remaining trajectories.

In our fixed-volume scenario for the economic evaluation of path forecasts, we focus on making decisions about when to sell the fixed amount of electricity.
The upper prediction band provides information on the highest probable price under a given SCP, while the lower prediction band reflects the lowest probable price.
For a risk-seeking decision, we may select the subperiod that achieves the highest value in the upper prediction band as the optimal time for selling. 
Conversely, for a risk-averse decision, we may select the subperiod time point with the highest value in the lower prediction band, thereby maximizing the lowest expected price.
We explore both choices and discuss their implications in more detail later.

\subsubsection{Benchmark trading strategies} \label{sssec:trading:naive}\label{ssec:trading:crystal}

\subsubsection*{Naive benchmarks}

We use three naive benchmark strategies that do not rely on any generated forecasts and instead execute trades at predefined times using the fixed volume of electricity.
In the Naive$_{\text{first}}$ benchmark, the energy producer submits a market order in period $t_0$, i.e., three hours before the delivery starts.
The Naive$_{\text{last}}$ benchmark proceeds by placing market orders in the last period $t_{10}$, i.e., 30 minutes before delivery, just before trading becomes restricted to selected zones.
Finally, the Naive$_{\text{avg}}$ benchmark distributes the total volume evenly across all the 10 periods $t_1,\ldots,t_{10}$, executing 10 equally sized trades.
It is worth noting that the Naive$_{\text{avg}}$ benchmark thus deviates from the previous assumption of a fixed-volume selling scenario where the total amount of available electricity is sold at a single subperiod, which might somewhat limit the fairness of the comparisons.

\subsubsection*{Crystal ball benchmarks and the realized trading potential}

The employed fixed-volume selling scenario has a maximum and minimum profit that can theoretically be achieved if the realizing observations of prices were known in advance.
Although this would obviously be impossible in any practical application, comparisons against the theoretical optimum might be of interest to assess the capabilities of the proposed forecasting models.
To that end, we construct a hypothetical crystal ball (CB) trading strategy, where we assume the future observations to be known and sell the available electricity during the subperiod with the highest (for the maximum profit CB benchmark) or the lowest (for the minimum profit CB benchmark) realized price.
We denote the profits from these two benchmarks by CB$_{\text{max}}$ and CB$_{\text{min}}$, respectively.
We can then define the realized trading potential (RTP) of a given combination of a forecasting model $A$ and a trading strategy as 
\begin{equation*}
\text{RTP}_A = \frac{\text{Profit}_A - \text{CB}_{\text{min}} }{\text{CB}_{\text{max}} - \text{CB}_{\text{min}}} \times 100,
\end{equation*}
where $\text{Profit}_A$ is the sum of the trading strategy's profits over the entire 200-day test period when using the predictions of model $A$.
The RTP, which can take values from 0 to 100, can be interpreted as the fraction of the maximum profit that can be achieved (times 100).

\subsubsection{Tailoring the conditional generative model to optimize trading profits}\label{ssec:cgm:newloss}

As discussed in Section \ref{ssec:CGM}, the CGM can be trained with alternative loss functions.
Here, we investigate an adaptation to potentially improve the economic aspects of the CGM model predictions, by combining the previously introduced economic evaluation as a custom loss function with energy score.
Specifically, we employ the majority-vote strategy for the fixed-volume trading scenario and integrate an additional loss component that measures the difference between the optimum index derived from the generated path trajectories and the observed optimal subperiod index obtained from the realizing ID prices.

Following Eq.\ (\ref{eq:majority-vote}), we derive the index $\tilde{J}_{d, h}$ of the optimal subperiod for selling based on the path trajectories $\{\tilde{\bm{X}}^m_{d, h}\}_{m=1}^M$ generated by the CGM by applying the majority vote strategy.
Let $J^{\text{obs}}_{d, h}$ denote the index of the subperiod with the highest observed ID price,
\begin{equation*}
    J_{d,h}^{\text{obs}} = \argmax_{j \in \{1,\ldots,10\}} X_{d,h,t_j}.
\end{equation*}
The custom loss function for the CGM is then defined as
\begin{equation*}
    \ell_{d, h} = (1 - \omega) \cdot \frac{1}{2} \cdot \text{ES}_{d, h} + \omega \cdot \left( \frac{1}{100} \cdot \big( \tilde{J}_{d, h} - J^{\text{obs}}_{d, h} \big)^2 \right),
\end{equation*}
where $\omega$ controls the weight of each component.
The ES component is divided by two to ensure a comparable magnitude of typical values encountered during the model optimization.

In the next section, we present results for $\omega = 0.5$, as preliminary experiments suggest that an equally weighted loss results in a better trade-off between statistical and economic performance\footnote{Results are available from the authors upon request.}.
The CGM approach trained solely on the energy score is denoted as "CGM (ES loss)", and the one trained with the custom loss that integrates the economic evaluation measure is denoted as "CGM (custom loss)".
The performances of both CGM variants are investigated.

\section{Results}\label{sec:results}

In this section, we present the results of both the statistical and economic evaluation of the generated path forecasts of the CGM and the statistical benchmark models.

\subsection{Statistical evaluation}

\begin{figure}
    \centering
    \includegraphics[width = 0.49\textwidth]{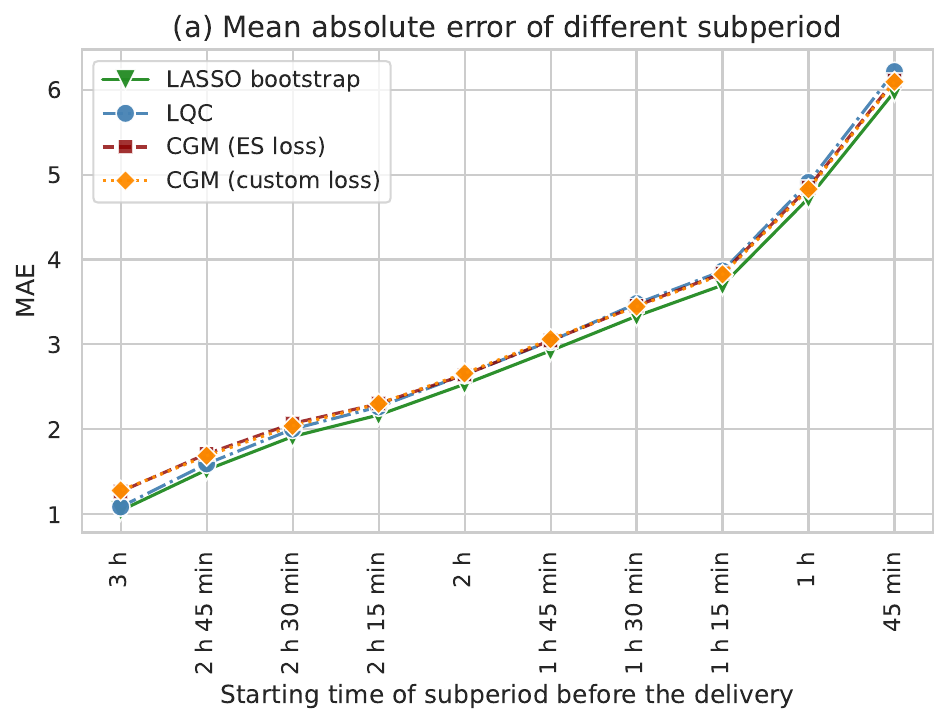}
    \includegraphics[width = 0.49\textwidth]{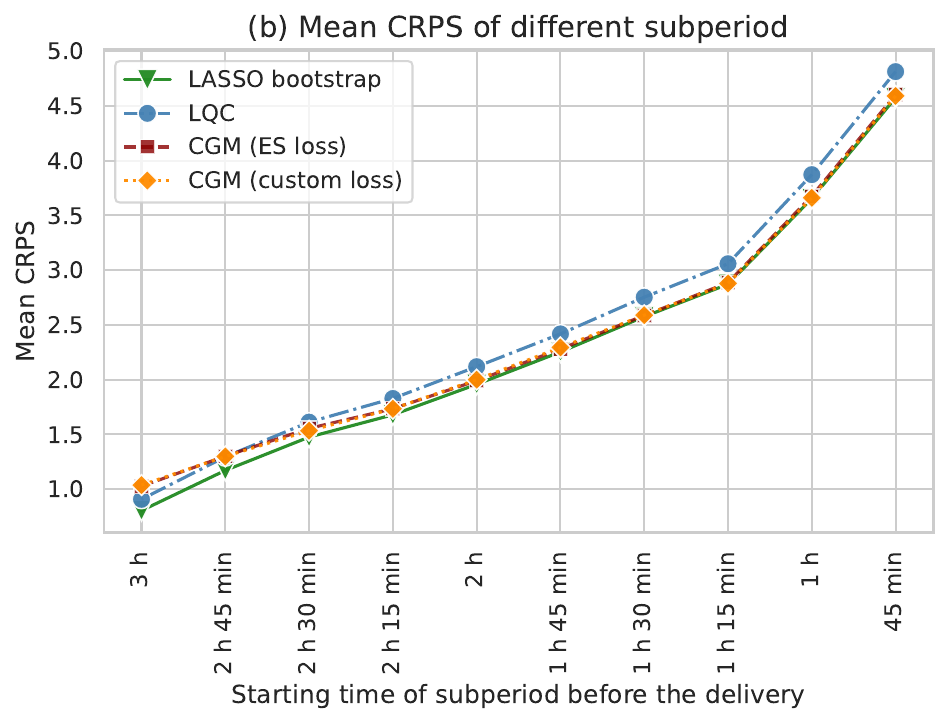}
    \caption{Mean absolute error (a) and CRPS (b) of different forecasting methods for each subperiod (margin) of the ID price path.}
    \label{fig:univ_score} 
\end{figure}

\begin{figure}
    \centering
    \includegraphics[width = \textwidth]{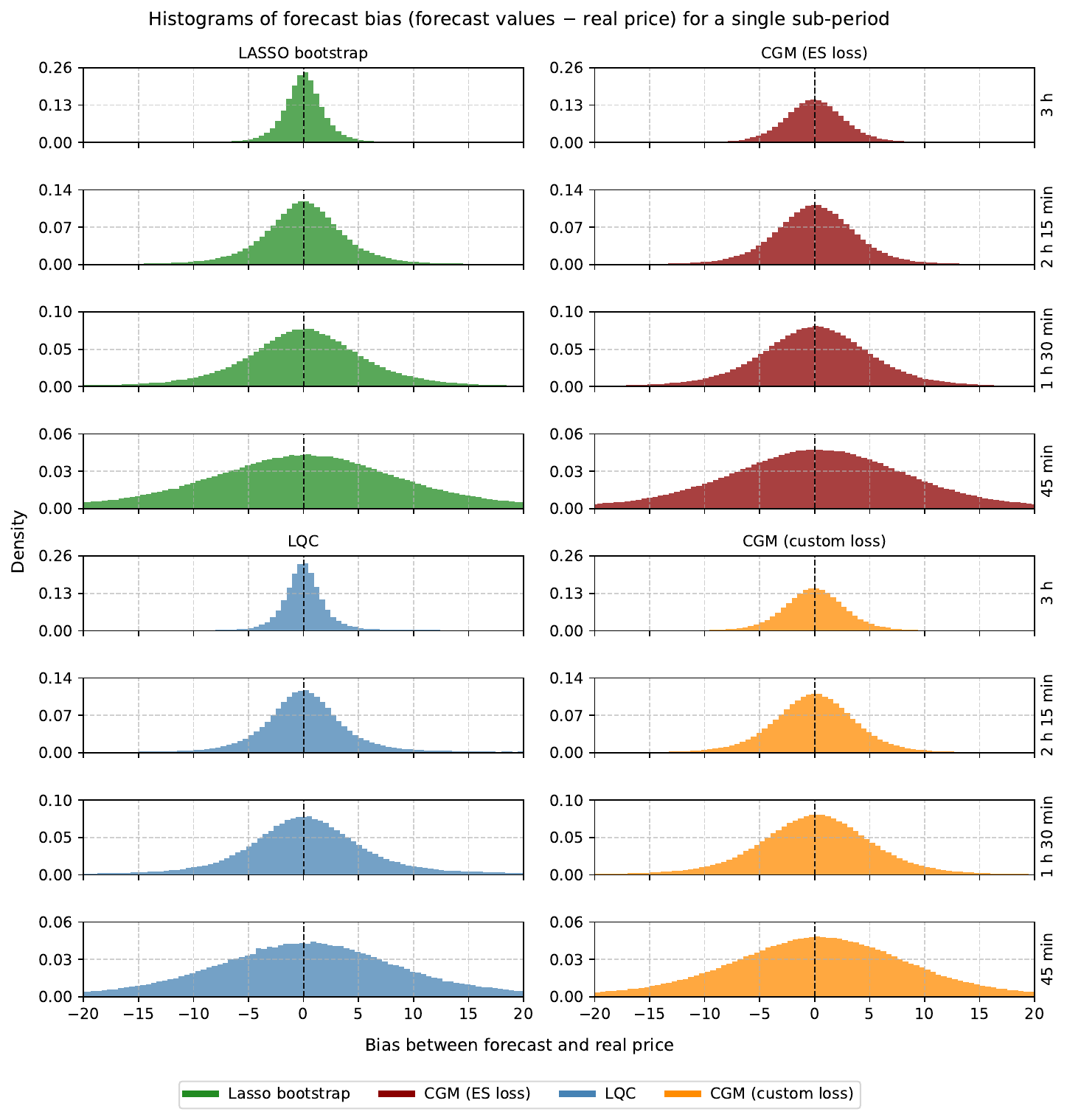}
    \caption{Distribution of biases (computed as forecast sample minus observed ID price) for each subperiod of the ID price path, over all hourly markets in the 200-day test period.}
    \label{fig:univ_dist} 
\end{figure}

We first compare the univariate performance of the different methods at each margin in terms of the mean absolute error of the median forecast and the CRPS in Figure \ref{fig:univ_score}.
Both evaluation metrics increase as the time subperiod approaches the target delivery time, indicating that it is harder to make accurate forecasts closer to delivery.
Figure \ref{fig:univ_score}(a) shows the absolute error, averaged over all hourly markets in the 200-day test dataset for each subperiod.
The LASSO bootstrap benchmark consistently outperforms other approaches across all subperiods closely followed by the LQC benchmark and two CGM variants.
In terms of CRPS shown in Figure \ref{fig:univ_score}(b), the LASSO bootstrap benchmark also performs best among all methods, especially for subperiods further from the delivery time.
For subperiods closer to the target delivery, the two CGM variants show comparable performance to the LASSO bootstrap benchmark.
The two considered CGM variants consistently show almost no difference in performance over both metrics.

To further investigate the distribution of the forecast errors of these different approaches, we present histograms of the differences between each marginal forecast sample and the observed ID price in Figure \ref{fig:univ_dist}. 
These histograms are shown for four selected time subperiods.
In the first subperiod, which begins three hours before the target delivery, we observe that the bias distributions of the forecasts from two CGM variants are notably wider compared to the two statistical benchmarks.
Conversely, in the last subperiod, starting 45 minutes before delivery, the two CGM variants produce slightly narrower bias distributions than the benchmarks.
The increasing variance of bias distributions from the first to the last subperiod further underscores the increased complexity of estimating ID prices as the delivery time approaches.

\begin{table}
  \begin{center}
    \small
      \caption{Mean values of multivariate proper scoring rules of the price path forecasts generated by the LASSO bootstrap and the LQC benchmark, and the two CGM variants. The results are shown separately for on- and off-peak hours. The best scores in each column are highlighted in bold.}
    \label{tab:statistical}
    \begin{tabular}{lcccccccc} 
    \hline
       \multicolumn{1}{c}{} & \multicolumn{2}{c}{ES} & \multicolumn{2}{c}{DSS} & \multicolumn{2}{c}{VS-1} & \multicolumn{2}{c}{VS-0.5}\\
      \cline{2-9}
      & on-peak & off-peak & on-peak & off-peak & on-peak & off-peak & on-peak & off-peak \\
      \hline
      LASSO bootstrap & \textbf{10.18} & \textbf{8.04} & 38.43 & 30.63 & 34.06 & \textbf{29.98} & \textbf{0.70} & \textbf{0.58} \\
      LQC & 10.86 & 8.86 & 45.10 & 39.05 & 41.23 & 34.96 & 0.83 & 0.73 \\
      CGM (ES loss) & 10.30 & 8.20 & 34.76 & 29.82 & \textbf{33.46} & 30.43 & \textbf{0.70} & 0.64 \\
      CGM (custom loss) & 10.37 & 8.20 & \textbf{34.41} & \textbf{29.60} & 33.53 & 30.33 & \textbf{0.70} & 0.63 \\
      \hline
    \end{tabular}
  \end{center}
\end{table}

Next, we compare the multivariate performance of the different methods.
Table \ref{tab:statistical} shows results for the last 200 days of the out-of-sample period in terms of the mean energy score, the Dawid-Sebastiani score, and two variants of the variogram score (VS-1, VS-0.5).
The evaluation is divided into on-peak hours (8:00-19:00) and off-peak hours (the remaining 12 hours of the day).

In general, different evaluation metrics suggest different best-performing methods, and no single method consistently outperforms the others across all metrics.
However, the LQC benchmark exhibits the weakest performance throughout, in particular for the DSS and VS\footnote{The relative performance of the LQC benchmark differs from the results in \citet{ser:mar:wer:22} due to a mistake in their code for preprocessing the data.}.
The other three methods are generally comparable, aligning with the univariate evaluation results.
The LASSO bootstrap benchmark performs best in terms of the ES, but is outperformed by the two CGM variants when evaluated using VS-1, VS-0.5 during on-peak hours, and in terms the DSS. 
These results suggest that the CGM variants are better in capturing the temporal dependence structure of the price paths, particularly during on-peak hours, which are more critical periods for trading markets, whereas the LASSO bootstrap benchmark exhibits a smaller bias.
That said, the score differences between the CGM models and the LASSO bootstrap benchmark tend to be mostly minor. 

\begin{figure}[tb]
    \centering                   
    \includegraphics[width = .49\textwidth]{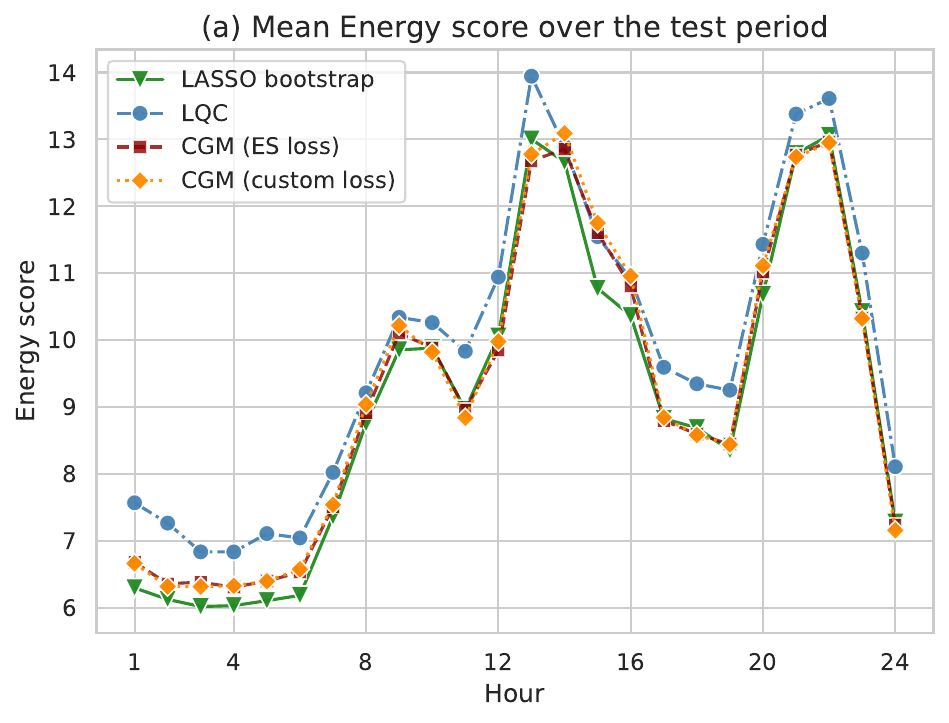}           
    \includegraphics[width = .49\textwidth]{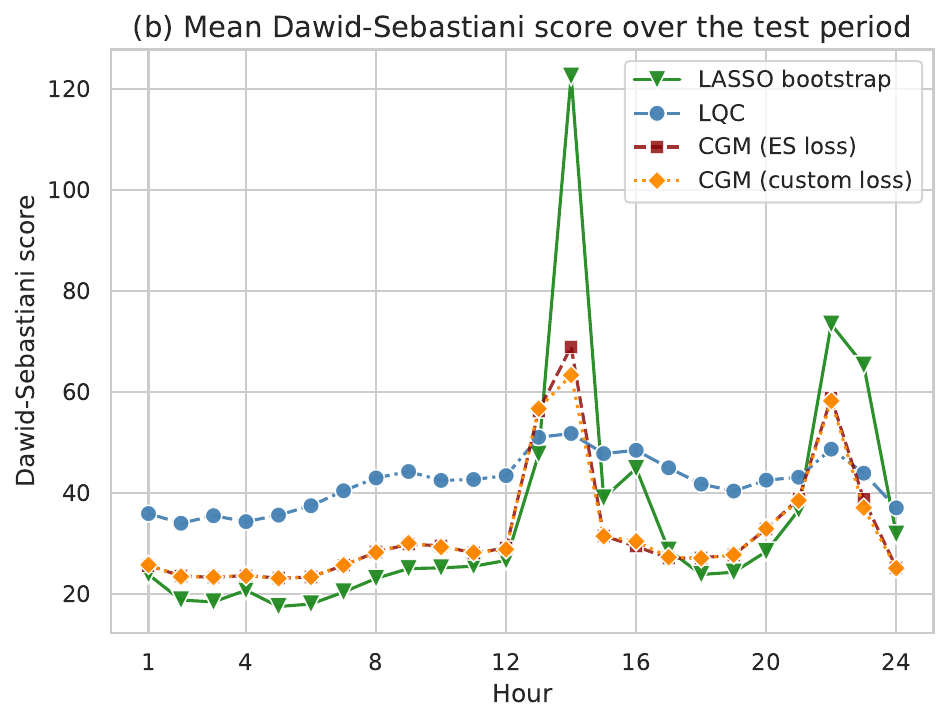}            
    \includegraphics[width = .49\textwidth]{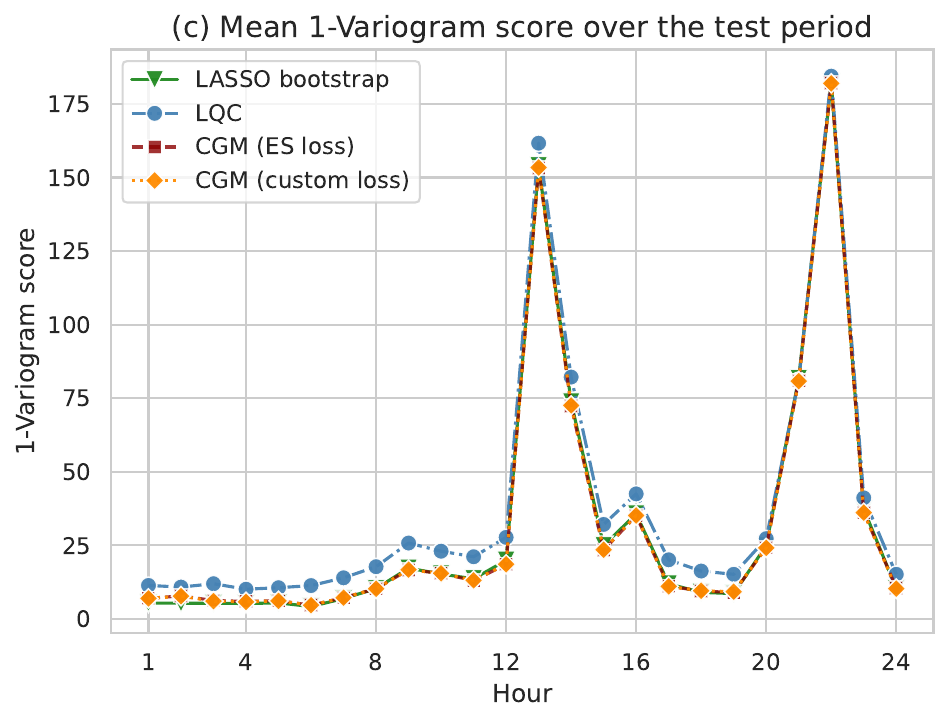}           
    \includegraphics[width = .49\textwidth]{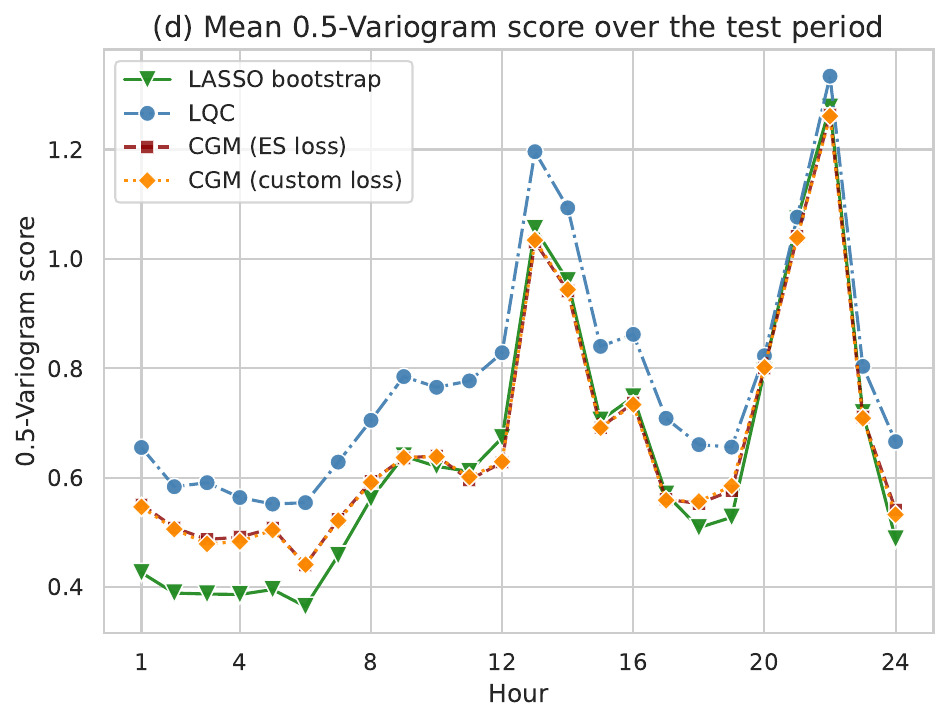}
    \caption{Mean values of the different multivariate proper scoring rules over 200 days in the test period for each hourly market.}
    \label{fig:scores} 
\end{figure}

Figure \ref{fig:scores} shows the mean value of the different scoring rules for each hourly market.
Across all four panels, two primary peaks are evident, roughly corresponding to 11:00–15:00 and 20:00–23:00, which likely align with high electricity demand during midday and evening hours.
During these peak periods, the CGM variants outperform the LASSO bootstrap benchmark in terms of DSS and both VS metrics, while the LASSO bootstrap method achieves better results in terms of the ES.
During the remaining periods with lower score values, the LASSO bootstrap method generally performs better than all other methods.
The LQC benchmark consistently performs worst over almost all hourly markets, with few exceptions. 
During the peak periods, which are typically more critical for real-world trading decisions, the CGM variants exhibit comparatively stronger performance, particularly in capturing temporal dependencies within the forecast trajectories.
In contrast, the LASSO bootstrap method shows notable outliers with by far the worst DSS values observed across all methods during periods with the highest overall forecast errors.

\subsection{Economic evaluation based on trading profits}\label{ssec:profits}

\begin{figure}
    \centering                   
    \includegraphics[width = .49\textwidth]{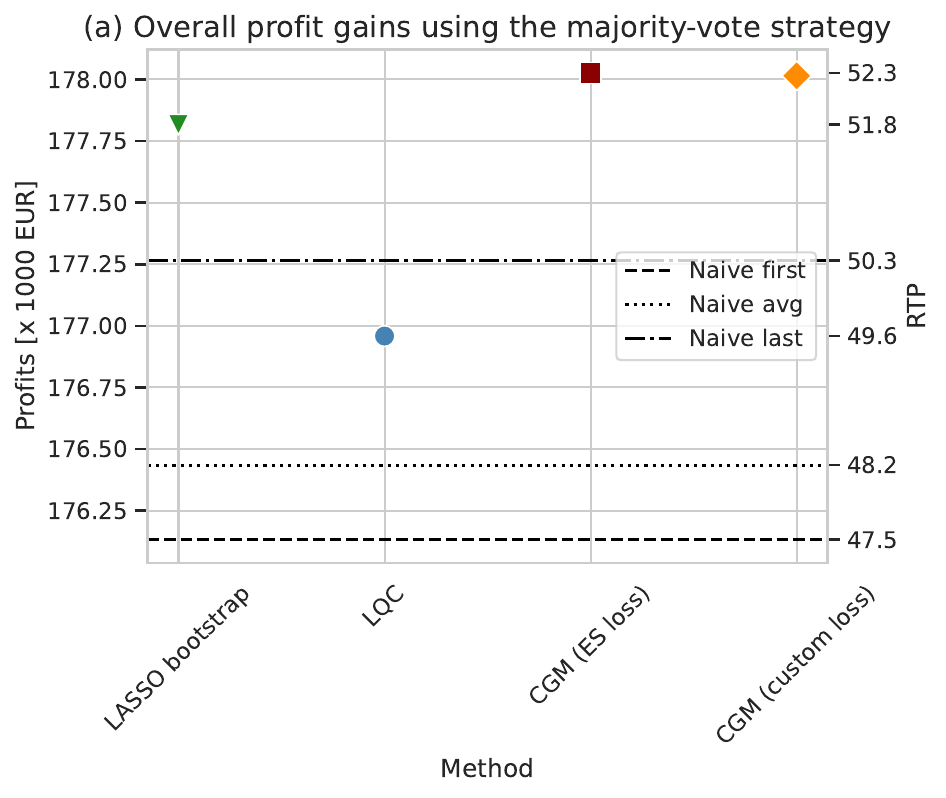}      
    \includegraphics[width = .49\textwidth]{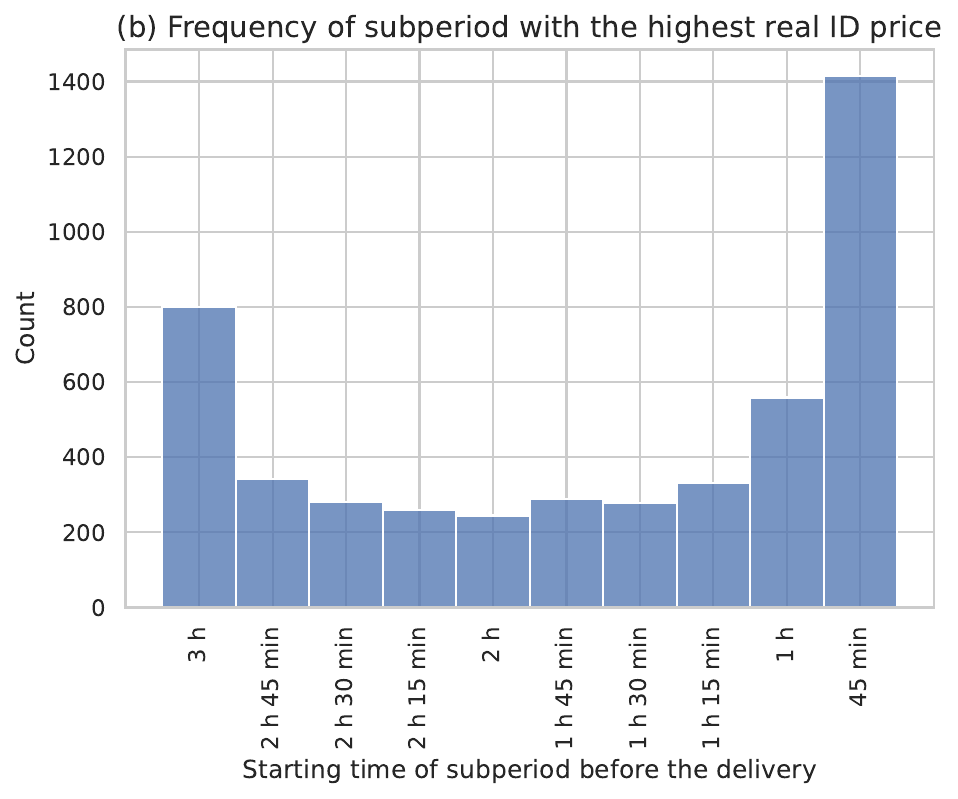}
    \caption{(a): Overall trading profit gains using the majority-vote strategy described in Section \ref{ssec:trading:band} in terms of the nominal profit (left axis) and the realized trading potential (right axis; see Section \ref{ssec:trading:crystal}). (b): Distribution of the subperiod with the highest observed price during the 10 considered subperiods within the test set.}
    \label{fig:profits_majority} 
\end{figure}

We follow the fixed-volume selling scenario introduced in Section \ref{ssec:Economic:evaluation}, which aims at maximizing the profit from selling 1~MWh in each hourly load period during the 200-day test period, based on the majority vote and prediction band strategies.
The trading profits based on the observed ID price at the selected time are computed as the sum of the profits for all hourly markets over the 200-day test period.

Figure \ref{fig:profits_majority}(a) compares the total profit gains of the different path forecasting methods using the majority-vote strategy.
For reference, the results from three naive benchmarks presented in Section \ref{sssec:trading:naive} are included as baselines. 
Among the four forecasting methods, the CGM trained with the ES performs best, closely followed by the CGM trained with the custom loss function. 
This is somewhat unexpected as the custom loss function was explicitly designed to incorporate economic evaluation during the model training. 
The LASSO bootstrap benchmark outperforms the LQC method and all three naive benchmarks, and is only slightly worse than the two CGM variants.
Further, the relative differences in the overall trading profits remain relatively small across all considered methods and benchmarks.

The realized trading potential indicated that the best naive strategy, Naive$_{\text{last}}$, achieves an RTP of 50.3.
In comparison, the CGM variants yield RTP values of approximately 52.3, representing a 4\% improvement over Naive$_{\text{last}}$.
The LASSO bootstrap benchmark, with an RTP of 51.8, provides a 3\% improvement.
Relative to Naive$_{\text{first}}$ with an RTP of 47.5, corresponding to a simple market sell order submitted 3 hours before delivery, the CGM variants achieve a 10\% improvement, while the LASSO bootstrap shows a 9\% improvement.

Interestingly, the Naive$_{\text{last}}$ benchmark performs well, even outperforming the LQC benchmark.
To investigate this further, we analyzed the observed prices within considered subperiods over the test period.
Figure \ref{fig:profits_majority}(b) shows the frequency of indices with the highest observed ID price for all hourly markets, indicating the highest price for a given hourly market most frequently occurs during the last subperiod (i.e., from 45 to 30 minutes before delivery).
This pattern explains the strong performance of Naive$_{\text{last}}$ relative to other naive benchmarks.

\begin{figure}
    \centering                   
    \includegraphics[width = .9\textwidth]{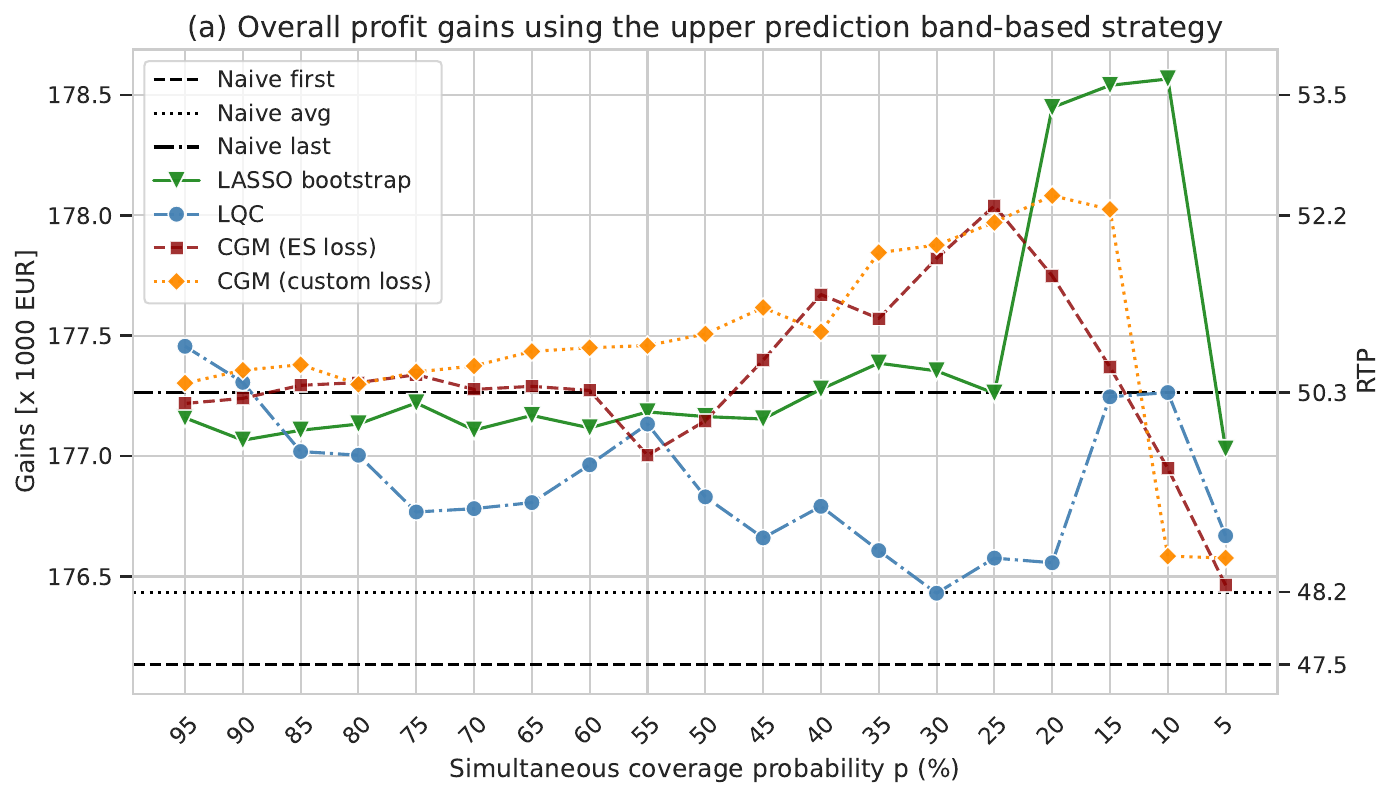}
    \includegraphics[width = .9\textwidth]{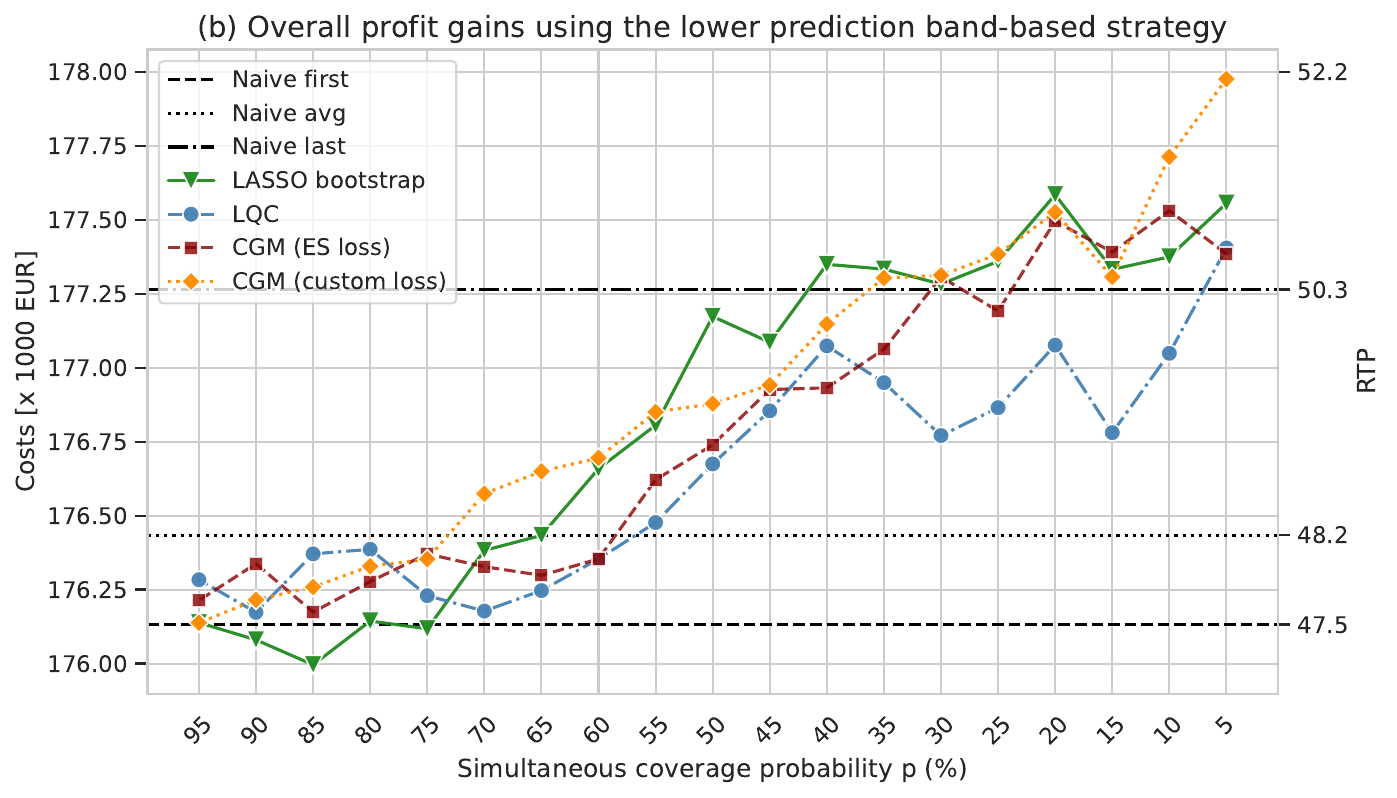}
    \caption{Overall trading profit gains using the prediction band-based strategy based on the upper prediction band (a) and the lower prediction band (b), in terms of the nominal profit (left axis) and the realized trading potential (right axis).}
    \label{fig:profits_band} 
\end{figure}

As discussed in Section \ref{ssec:trading:band}, prediction bands can be derived from a collection of path forecasts. 
For evaluation and comparison, we first need to specify the simultaneous coverage probability.
Here, SCP values ranging from 5\% to 95\% are considered as ex-post selected thresholds for a more generalized analysis. 
In real-time trading, the optimal SCP value leading to the highest profits varies over time and of course needs to be selected ex-ante, for example based on historical data, as suggested in \citet{ser:mar:wer:22}. 

Figure \ref{fig:profits_band} illustrates the profit gains achieved using the prediction band-based strategy, based on both the upper and lower prediction bands with selected SCP values. 
As discussed in Section \ref{ssec:trading:band}, the decision to determine the optimal selling time based on either the upper or lower prediction band reflects the trader's risk preference.
Our observations indicate that no forecasting method consistently outperforms the others across all SCP values in both cases. 
However, the profit gains associated with the upper prediction band are generally higher than those from the lower prediction band, suggesting that taking on a relatively higher level of risk may yield better returns.

In the lower prediction band-related results shown in Figure \ref{fig:profits_band}(b), there is a clear trend of increasing profits as the SCP decreases.
In contrast, the upper prediction band-related results shown in Figure \ref{fig:profits_band}(a) do not exhibit a clear trend.
At very low SCP values, the remaining path trajectories for deriving lower prediction bands are those with consistently high predicted ID prices, while for the upper prediction bands, the remaining path trajectories correspond to those with consistently low predicted ID prices.
This makes the results of profit gains more diverse, as observed for both types of bands in the illustrations, compared with other SCP values.

Focusing on the upper prediction band results that yield higher profits in Figure \ref{fig:profits_band}(a), we observe that within the middle SCP range (25\%–75\%), which is more commonly used, the CGM trained with the custom loss function consistently achieves the best performance. 
This aligns with the intended purpose of the custom loss design.
The CGM trained with the ES closely follows, with both CGM variants outperforming all benchmark methods. 
Although the LASSO bootstrap approach performs reasonably well, its performance under this strategy is not as strong as in the majority vote strategy, and it fails to outperform the best naive baseline. The LQC method performs worst throughout.

\section{Conclusions}\label{sec:conclusions}

We propose a new approach for electricity price forecasting in continuous intraday markets, utilizing conditional generative machine learning models to produce probabilistic path forecasts. 
The proposed CGM approach generates multivariate path trajectories directly as the output of a generative neural network, trained using the energy score, which is a mathematically principled loss function for multivariate probabilistic forecasts. 
A key advantage of this approach is the ability to bypass the separate modeling of marginal distributions and temporal dependencies, which is the cornerstone of many alternative multivariate forecasting approaches. 
By conditioning on exogenous input variables, such as wind and load data, CGMs can flexibly incorporate information from additional predictors in both the marginal distributions as well as the temporal dependencies. 
Further, the CGMs can be trained with custom loss functions, for example aiming to integrate specific economic objectives related to trading profits in electricity markets.

An important aspect for evaluating multivariate EPF models is to not only apply commonly used statistical evaluation metrics in the form of suitable multivariate proper scoring rules, but also to evaluate the forecasts from a practically oriented, economic perspective.
To that end, we proposed two tailored trading strategies based on multivariate probabilistic information, the majority vote strategy and the prediction band-based strategy, to evaluate the economic performance of path forecasts in a fixed-volume selling scenario.
The results show that while no single model consistently outperforms all others across all statistical and economic evaluation metrics, the CGM framework demonstrates good performance in both aspects compared to two state-of-the-art statistical benchmark methods. 
Specifically, the CGM is better able to capture temporal dependencies, particularly during peak electricity usage hours. 
In terms of the economic evaluations, a naive benchmark approach of placing sell orders always at the last time subperiod performed well due to the typical trends of observed ID price paths. 
Nevertheless, CGMs improved profit gains over this benchmark by 4\% in the majority-vote strategy, and yield the highest overall trading profits across all considered approaches.
In the prediction band-based strategy, CGMs showed clear advantages, particularly when trained with a custom loss that integrates economic objectives, further highlighting their potential benefits for trading scenarios.

To the best of our knowledge, our work is the first to introduce generative machine learning methods for forecasting ID electricity price paths.
A promising avenue for future work lies in advancing economic evaluation methodologies. 
Realistic trading scenarios provide valuable insights into model performance from a decision-maker’s perspective, serving as a practical complement to traditional statistical metrics.
By bridging the gap between forecasting accuracy and economic impact, this study contributes to the literature on the economic evaluation of forecasts \citep{mac:uni:wer:23,yar:pet:21}. 
Moving beyond the fixed-volume selling scenario explored here, it would be interesting to investigate other scenarios in realistic trading markets.
Some of the ideas for making trading decisions proposed in this study could be adapted and may also require substantial modifications for certain scenarios, underscoring the need for further research in this domain.
Beyond the specific trading scenario, the optimal use of multivariate probabilistic forecasts in deriving optimal trading strategies represents another interesting topic for future research.

Furthermore, while our attempts to integrate the economic aspects into the loss function for training the generative models showed some promise in terms of the realized trading profits, the overall improvements over naive benchmark strategies remain limited. 
From a methodological perspective, it would be interesting to further investigate the role of the loss function in training generative models for multivariate probabilistic forecasting, with possible choices including a plethora not only of available multivariate proper scoring rules \citep{pacchiardi_etal_2024_probabilistic}, but also of potential ways to incorporate economic aspects.

\section*{Acknowledgments}
The authors would like to thank Grzegorz Marcjasz for comments on an earlier version of the manuscript.
The work of J.C.\ and S.L.\ was funded by the Vector Stiftung within the Young Investigator Group ``Artificial Intelligence for Probabilistic Weather Forecasting''. In addition, J.C.\ has been funded by by the German Research Foundation (DFG) through project T4 ``Development of a deep learning prototype for operational probabilistic wind gust forecasting'' of the Transregional Collaborative Research Center SFB/TRR 165 ``Waves to Weather''.
S.L.\ and M.S.\ acknowledge support of the Klaus Tschira Foundation.
Further, this work was partially supported by the Ministry of Science and Higher Education (MNiSW, Poland) through Diamond Grant No.\ 0009/DIA/2020/49 (to T.S.) and the National Science Center (NCN, Poland) through grant No.\ 2018/30/A/HS4/00444 (to R.W.).

\bibliographystyle{myims2}
\bibliography{ref}

\end{document}